\documentclass{cicirello}
\pdfoutput=1

\usepackage{graphicx}
\usepackage{natbib}
\usepackage{doi}

\title{Design, Configuration, Implementation, and Performance of a Simple 32 Core Raspberry Pi Cluster}
\author{Vincent A. Cicirello}
\orcid{0000-0003-1072-8559}
\reportnum{DIST-17-004}
\reporturl{https://reports.cicirello.org/17/004/VAC-TR-17-004.pdf}
\copyrightyear{2017}
\date{August 2017}

\address{Computer Science\\
Stockton University\\
Galloway, NJ 08205 USA\\
\url{https://www.cicirello.org/}
}

\ACM{D.1.3; G.4; I.1.2}

\subject{In this report, I describe the design and implementation of an inexpensive, eight node, 32 core, cluster of
raspberry pi single board computers, as well as the performance of this cluster on two computational tasks,
one that requires significant data transfer relative to computational time requirements, and one that does not.
We have two use-cases for the cluster: (a) as an educational tool for classroom usage, such as covering
parallel algorithms in an algorithms course; and (b) as a test system for use during the development of 
parallel metaheuristics, essentially serving as a personal desktop parallel computing cluster.
Our preliminary results show that the slow 100 Mbps networking of the raspberry pi significantly limits 
such clusters to parallel computational tasks that are either long running relative to data communications
requirements, or that which requires very little internode communications.  Additionally, although the raspberry pi 3 
has a quad-core processor, parallel speedup degrades during attempts to utilize all four cores of all cluster nodes
for a parallel computation, likely due to resource contention with operating system level processes. However, 
distributing a task across three cores of each cluster node does enable linear (or near linear) speedup.}

\abstract{In this report, I describe the design and implementation of an inexpensive, eight node, 32 core, cluster of
raspberry pi single board computers, as well as the performance of this cluster on two computational tasks,
one that requires significant data transfer relative to computational time requirements, and one that does not.
We have two use-cases for the cluster: (a) as an educational tool for classroom usage, such as covering
parallel algorithms in an algorithms course; and (b) as a test system for use during the development of 
parallel metaheuristics, essentially serving as a personal desktop parallel computing cluster.
Our preliminary results show that the slow 100 Mbps networking of the raspberry pi significantly limits 
such clusters to parallel computational tasks that are either long running relative to data communications
requirements, or that which requires very little internode communications.  Additionally, although the raspberry pi 3 
has a quad-core processor, parallel speedup degrades during attempts to utilize all four cores of all cluster nodes
for a parallel computation, likely due to resource contention with operating system level processes. However, 
distributing a task across three cores of each cluster node does enable linear (or near linear) speedup.}

\begin{document}

\maketitle

\section{Introduction}

The concept of cluster computing is now well over 30 years old. 
Back in 1994, researchers at NASA developed a much less expensive 
alternative to traditional supercomputers~\citep{BeowulfNasa1995}.
Rather than the expense of a specially designed supercomputer, they 
built the equivalent using 16 off-the-shelf standard desktop PCs, 
networked together with a specially designed operating system, 
along with parallel computing APIs, such that it operates as a 
single parallel system.  They named their system Beowulf, 
inspired by the epic poem of the same name, which describes Beowulf 
as having the strength of 30 men: ``that he thirty men's grapple has in his hand, 
the hero-in-battle''~\citep{BeowulfEpic}.

Though it began as the name of one specific parallel 
computing system, today ``Beowulf cluster'' more 
generally refers to any parallel computing system built 
from off-the-shelf computers networked together to 
operate as one.  With the growing availability of 
inexpensive single board computers, such as the Raspberry Pi,
Beowulf style clusters are proliferating, especially 
in educational environments~\citep{Adams:2016,Adams:2015}.  
The very low cost of such single board computers make it possible
for more people to explore cluster computing concepts, as it 
is possible to build an eight node compute cluster for
less than the cost of a desktop PC.  The low cost of building 
such systems is enabling more in depth coverage
of parallel and distributed computing at the undergraduate 
course level~\citep{Adams:2016,Adams:2015,Matthews:2016}.

In this report, I describe the design and implementation 
of an inexpensive, eight node, 32 core, cluster of
raspberry pi single board computers, as well as the 
performance of this cluster on two computational tasks,
one that requires significant data transfer relative 
to computational time requirements, and one that does not.
Our cluster is designed to be small, lightweight, and 
easily transportable.  Its intended purpose is two-fold:
(a) as an educational tool for classroom use, and (b) as 
a test system for my research in parallel metaheuristics.
First, the cluster will be used in a classroom setting, 
such as in an algorithms course for coverage of parallel 
algorithms.  As such, it must be easy to transport to 
classrooms, with minimal setup once there, 
such as on a utility cart with integrated power.
We design the cluster from eight of the single board 
computer Raspberry Pi 3. The Raspberry Pi was developed 
originally as an inexpensive way to enable more children to experience 
computer science education at an earlier 
age---eliminating the expense of purchasing a computer 
(a single Raspberry Pi 3 costs \$35 plus another \$35 for 
a case, power supply, and memory card)~\citep{RaspPi}.  
The Raspberry Pi 3 has a quad-core ARM processor at 1.2 GHz, 
which can be overclocked to 1.6 GHz, although we have not done so.

Our second use-case for the cluster is as a test system for 
parallel metaheuristics, such as our research on parallel
simulated annealing~\citep{cicirello2017SoCS2}.  Although the 
CPUs are slower than we might use in a production system,
it enables easily experimenting with algorithm design concepts 
in an environment that is not impacted by other system users
(e.g., using it as a ``desktop'' personal parallel compute cluster). 

This report is organized as follows. Section~\ref{sec:design} 
provides details of our cluster, including hardware requirements and cost,
and system configuration including networking.
In Section~\ref{sec:perform} we present the results of some experiments 
on two computational tasks. We implement our experiments using Java, and 
Java RMI for communicating among the cluster nodes.  All of our source code is
open source, and is available via 
GitHub~\citep{Cicirello:GitHub:2017}\footnote{\url{https://github.com/cicirello/ClusterPerformanceTests}}.  
The first task is parallel matrix multiplication (Section~\ref{sec:mult}), 
and the second is Monte Carlo estimation (Section~\ref{sec:pi}).  
Matrix multiplication requires distributing substantial 
data, relative to the computational time 
needed for the multiplication; while Monte Carlo 
estimation requires very little internode communications.  
There are also numerous parallel matrix multiplication algorithms 
available (e.g.,~\citep{Karstadt:2017,Ballard:2012,Chatterjee:1999}), 
and parallel matrix multiplication is a topic commonly covered in 
algorithms courses and textbooks (e.g.,~\citep{cormen2009introduction}).
For Monte Carlo integration, we implement the obvious and 
straightforward parallel extension of the classic average value method (referred to
as ``method one'' in~\citep{Neal1993}).  

Our choice of Raspberry Pi as the single board computer limits our system 
to 100 Mbps Ethernet, so this pair of tasks was chosen to examine the effects of the Raspberry Pi's
slow networking (i.e., one task with substantial communication needs, and one with very little). 
We conclude with some observations and recommendations in Section~\ref{sec:conclude}.

\section{Cluster Design and Implementation}\label{sec:design}

\subsection{Hardware}

Table~\ref{tab:equipment} provides a summary of the equipment and supplies required,
along with costs.  The total cost of the cluster is approximately \$939.  These costs
include some items that are not strictly required for the cluster, which I will identify in what follows,
that are included here to enable convenient use of the raspberry pis for other purposes separate 
from the cluster.

\begin{table}[t]
\centering
\caption{Equipment and cost summary}
\begin{tabular}{l|rr|r} \hline
Hardware					& Quantity	& Item Cost	& Line Cost \\ \hline
Raspberry Pi 3 (bundle including heatsinks,	& 8			& \$69		& \$552	\\
16GB MicroSD card, power supply, HDMI cable, & & & \\
standard Raspberry Pi case)	& & & \\ \hline
Micro USB to USB Cable 		& 8			& \$3			& \$24	\\ 
(Tronsmart 20AWG charging cables) & & & \\ \hline
Cat5e Cable			& 9			& \$1			& \$9	\\ \hline
8-port network switch (TP-Link 8-Port 	& 1			& \$24			& \$24	\\
Gigabit Ethernet Desktop Switch) & & & \\ \hline
USB Ethernet Adapter 		& 1			& \$21		& \$21   \\
(StarTech USB 3.0 to Gigabit Ethernet) & & & \\\hline
Phone charger (Anker PowerPort 6)				& 2			& \$40			& \$80	\\\hline
Display (SunFounder 10.1 Inch HDMI 					& 1	& \$110	& \$110  \\
1280x800 HD IPS LCD) & & & \\ \hline
Logitech MK270 Wireless Keyboard and Mouse Combo & 1	& \$25	& \$25	\\\hline
SanDisk Cruzer Glide CZ60 128GB USB 2.0 Flash Drive	& 1	& \$25 & \$25 \\\hline
Surge protector / power strip	& 1		& \$11			& \$11	\\\hline
Utility cart with integrated power	& 1	& \$58			& \$58	\\ \hline
TOTAL	&	&	& \$939 \\\hline
\end{tabular}\label{tab:equipment}
\end{table}

\textbf{Raspberry Pi 3:}
Our cluster consists of eight Raspberry Pi 3 single board computers.
The Raspberry Pi 3 consists of a quad-core ARM 1.2 GHz CPU,
a Broadcom VideoCore IV GPU, 1GB RAM, 10/100 Mbps Ethernet, 2.4 GHz 802.11n wireless,
and Bluetooth.  Storage is via MicroSD.  The Raspberry Pi 3 has the following ports:
HDMI, 3.5mm audio/video jack, and four USB 2.0 ports.

We acquired bundles that include heatsinks and 16GB MicroSD 
cards, both of which we are using, and which
also included HDMI cables, and standard 5V 2.5A 
power supplies.  For the cluster, we only need 
one of the HDMI cables, and none of the power supplies.  
However, we wanted to have the power supplies available
to easily enable using the Pis of the cluster for 
other purposes (e.g., in a classroom setting).  Seven of the HDMI cables are
unnecessary, but the bundle price was less than buying everything unbundled.
The bundles also included standard Raspberry Pi 
cases. But again, we wanted to be able to conveniently
break the cluster down as needed to use for other purposes.  
Having the Pis in standard cases, rather than a cluster case,
makes this possible without having to physically unmount them from a rack.

\textbf{Storage:} The seven worker nodes are limited to the 16GB MicroSD card.
The master node additionally has a 128GB USB flash drive.

\textbf{Input / Output:}
In addition to accessing the cluster over our campus network, we've equipped the master node with 
a 10.1 inch HDMI LCD display, as well as a wireless keyboard/mouse combo.

\textbf{Networking:} The Raspberry Pi 3 is limited 
to 100 Mbps Ethernet.  We are using a gigabit switch
(TP-Link 8-Port Gigabit Ethernet Desktop Switch) 
simply for convenience (small size) and cost (it was cheap).
The master node handles network routing, and has a second 
Ethernet interface, StarTech USB 3.0 to Gigabit Ethernet,
for the connection to the Internet.  The Raspberry Pi 3 
has USB 2.0, and thus, it is not possible to get
gigabit rates out of this USB Ethernet adapter.  
USB 2.0 has a theoretical limitation of 480 Mbps,
but we are unlikely to get anywhere near that as the 
Pi's onboard Ethernet, and any other USB devices, 
share the single USB 2.0 bus.

\textbf{Power:} To minimize the number of power cords 
we need to plug in, we use two 
Anker PowerPort 6 phone chargers.  The Anker 
PowerPort 6 is a 5V, 12A charger, which supplies
a max current of 2.4A per port.  It has 6 ports, 
but we limit our use to 4 ports for 4 Raspberry Pis
to ensure each Pi has sufficient power, although 
it is unlikely that our Raspberry Pis will draw power at their
max specified rate of 2.5A.  We are using Tronsmart 20AWG 
Micro USB to USB charging cables.  It is especially important to use 
high quality heavy gauge cables, such as the 20AWG to 
ensure that the Raspberry Pis get the power that they need.
Thinner cables will not support the power draw of the Pi.

As will be indicated later, we have disabled 
Bluetooth and Wi-Fi, and only the master
node uses any USB devices (a second network 
interface, flash drive, and keyboard/mouse combo).
Using a USB amp meter, the master node running 
a compute intensive task that utilizes all four 
processor cores as well as Ethernet, with an 
HDMI display attached, as well as the second Ethernet adapter,
was observed to top out at approximately 0.8A.  We otherwise 
did not conduct extensive power experiments, as it does not 
look like we will approach the limitations of our power equipment.

Additionally, we keep the cluster on a utility cart 
with integrated power, to enable easily transporting to
classrooms as necessary (e.g., for classroom activities, 
such as related to parallel algorithms in an algorithms course). 

\subsection{Configuration}

We're using Raspbian as the operating system.  We obviously changed the password for user pi
to a more secure one than the default.  Additionally, we created a second user without sudo access
for executing all cluster applications.  
For hostnames, 
we used a standardized approach, specifically, $\{ rpi0, rpi1, \ldots, rpi7 \}$,
with $rpi0$ as the master node.  We also install iperf3 on all nodes (not installed by default),
which is useful for network testing.

\textbf{Disable Bluetooth and Wi-Fi:} We disable both Bluetooth and Wi-Fi on all nodes of the cluster
to conserve power.  This is accomplished by adding the following two lines to /boot/config.txt:
\begin{verbatim}
     dtoverlay=pi3-disable-bt
     dtoverlay=pi3-disable-wifi
\end{verbatim}

\textbf{Master Node Networking:} The master node handles network routing for the local cluster network.
We do not assign static IP addresses to the worker nodes.  Instead, we run a DHCP server and DNS server
on the master node (specifically dnsmasq, which can be installed on the master node with:
sudo apt-get install dnsmasq).  On the master node, 
we configure dnsmasq with the following (in /etc/dnsmasq.conf):
\begin{verbatim}
     interface=eth0
     listen-address=127.0.0.1
     dhcp-range=192.168.99.2,192.168.99.254,255.255.255.0,24h
\end{verbatim}
Note in the above that we use the master node's onboard Ethernet interface (eth0) for its connection
to the cluster's network.  We assign a static IP address, 192.168.99.1, to the master node (thus,
why the dhcp-range begins at 192.168.99.2).

\sloppy{Enable port forwarding on the master node, by uncommenting the following line
in the file /etc/sysctl.conf:}
\begin{verbatim}
     net.ipv4.ip_forward=1
\end{verbatim}

We also edit the master node's /etc/network/interfaces file to configure eth0 (the onboard Ethernet
for the local cluster) and eth1 (the USB Ethernet interface for the connection to the Internet) as 
follows:
\begin{verbatim}
     auto eth1
     allow-hotplug eth1
     iface eth1 inet manual
     post-up iptables-restore < /etc/iptables.ipv4.nat

     auto eth0
     allow-hotplug eth0
     iface eth0 inet static
     address 192.168.99.1
     netmask 255.255.255.0
     network 192.168.99.0
     broadcast 192.168.99.255
\end{verbatim}
\sloppy{Since we're assigning IP addresses dynamically to the worker nodes, we can simply use the default
/etc/network/interfaces on the workers.}

Although we don't anticipate the need for worker nodes to access the Internet during computational tasks,
they will need such access for OS and software updates.  We configure IP forwarding on the master node 
(via /etc/iptables.ipv4.nat) as follows:
\begin{footnotesize}\begin{verbatim}
     *filter
     :INPUT ACCEPT [0:0]
     :FORWARD ACCEPT [0:0]
     :OUTPUT ACCEPT [0:0]
     -A INPUT -i lo -j ACCEPT
     -A INPUT -i eth0 -j ACCEPT
     -A INPUT -i eth1 -m state --state RELATED,ESTABLISHED -j ACCEPT
     -A FORWARD -i eth1 -o eth0 -m state --state RELATED,ESTABLISHED -j ACCEPT
     -A FORWARD -i eth0 -o eth1 -j ACCEPT
     COMMIT
     *nat
     :PREROUTING ACCEPT [0:0]
     :INPUT ACCEPT [0:0]
     :OUTPUT ACCEPT [0:0]
     :POSTROUTING ACCEPT [0:0]
     -A POSTROUTING -o eth1 -j MASQUERADE
     COMMIT
\end{verbatim}\end{footnotesize}

\textbf{SSH Keys:}
We use ssh for a variety of reasons, including from within scripts on the master node to
start up processes on the worker nodes, as well as for administrative purposes such 
as shutting down or rebooting worker nodes.  Therefore, we generate ssh keys to simplify 
this, by executing the following on the master node for each of the two users (the default user
pi that has sudo access, and the user that we created for executing cluster applications):
\begin{verbatim}
     ssh-keygen -t rsa -C user@rpi0
\end{verbatim}
Note that we save the keys in the default location, and do not use a pass phrase.
Each of these keys must also be distributed to the seven worker nodes (rpi1, rpi2, etc) via the following:
\begin{footnotesize}\begin{verbatim}
     cat ~/.ssh/id_rsa.pub | ssh user@rpi1.local 'cat >> .ssh/authorized_keys'
\end{verbatim}\end{footnotesize}

\section{Performance}\label{sec:perform}

We examine the cluster performance on a couple simple, easily parallelized computational tasks.
The tasks are matrix multiplication (Section~\ref{sec:mult}) and Monte Carlo estimation of Pi 
(Section~\ref{sec:pi}).  There's a strange sense of irony computing Pi with a cluster of Pis.
Our code is implemented in Java using Java RMI for
remote execution on the cluster's worker nodes.  
All code is open source and is available via 
GitHub~\citep{Cicirello:GitHub:2017}\footnote{\url{https://github.com/cicirello/ClusterPerformanceTests}}.

\subsection{A Few Notes on Implementation}

Each worker node runs an RMI server.  The RMI servers are brought up at system startup, and remain
up until the system is either shutdown or rebooted.  The RMI servers running on the workers
provide the master node with Java methods for each parallel task.  These are then implemented
on the workers using multithreading.  The RMI server maintains a cached thread pool to minimize
the overhead associated with creating threads.  During RMI server startup, we also execute each
parallel algorithm once with four threads to: (a) initiate the thread pool with a few threads, and
(b) cause Java's JIT compiler to compile the most critical portions natively.
The parallel implementations on the master node utilizes one local thread for each worker node,
enabling concurrent RMI calls to the workers.

Complete implementation details can be found in the source code in the GitHub 
repository\footnote{\url{https://github.com/cicirello/ClusterPerformanceTests}}.

\subsection{Matrix Multiplication}\label{sec:mult}

Parallel matrix multiplication has been widely studied, and
is often used as one of the first examples of parallel algorithms in courses on
algorithms.  As such, there are many parallel algorithms 
available for matrix multiplication~\citep{Karstadt:2017,Ballard:2012,Chatterjee:1999}, many of which are
based on Strassen's method for matrix multiplication~\citep{Strassen:1969}.

In our experiments, we consider a restricted form of matrix multiplication, namely the
task of multiplying a matrix $M$ by a vector $V$:
$C = M * V$.  
This is easily implemented in parallel, by distributing the rows of $M$ among the 
available threads.  For example, consider $N$ threads: $\{ t_1, t_2, \ldots, t_N \}$.  
Each thread $t_i$ is assigned a submatrix $M_i$ of $M$ consisting of $R/N$ of the rows of $M$, where $R$ is the number
of rows of matrix $M$.  If $R$ is not divisible by $N$, then some threads will have one more row than the others.
Thread $t_i$ computes $C_i = M_i * V$, which is $R/N$ of the rows of $C$.  $C$ is then formed by combining all of the
$C_i$.

Results using a single Raspberry Pi, for a matrix $M$ with 3000 rows and 3000 columns, are summarized in 
Table~\ref{tab:mult}, for one to four concurrent threads.  For this first batch of experiments, all threads run locally
on the master node (Java RMI is not used).
The times are in seconds, and averaged across 10 runs.  Speedup is relative to the single thread time.
For up to three threads, speedup is approximately linear.  At four threads, we are maximizing processor core usage,
and speedup degrades with very little gain over three thread performance.  This likely relates to OS processes, etc,
getting time on the processor.

\begin{table}[t]
\centering
\caption{Timing results for matrix multiplication on single Raspberry Pi.}\label{tab:mult}
\begin{tabular}{lrr}\hline
			&	Time (seconds)	& Speedup \\\hline
1 thread	&	0.196	& 1.00 \\
2 threads	&	0.103	& 1.89 \\
3 threads	&	0.067	& 2.90 \\
4 threads	&	0.062	& 3.17 \\\hline
\end{tabular}
\end{table}

We did not conduct a broader experiment with matrix multiplication across the nodes of the cluster for the following reason.
The Raspberry Pi 3 is limited to 100 Mbps networking.  Our matrix $M$ is implemented as a 2D array of doubles in Java
(double precision floating point).  Although only part of $M$ must be transmitted to each worker node, and each worker node only sends
its portion of $C$ back to the master node, overall during the task the entire matrices $M$ and $C$ are transmitted.
The time, $T$, required to transmit these matrices, assuming the ideal case that we get the full 100 Mbps is as follows:
\begin{equation}
T = 2 \text{ matrices} * \frac{9 * 10^6 \text{ doubles}}{\text{matrix}} * \frac{64 \textrm{ bits}}{\text{double}} * \frac{1 \text{ second}}{10^8 \text{ bits}} = 11.52 \text{ seconds}
\end{equation}
The entire task if computed on a single core of the local machine only requires approximately 0.2 seconds on average.  We clearly cannot afford the
cost in time of distributing the matrix across the nodes of the cluster relative to the time to simply compute the solution locally. 
The 100 Mbps networking of the Raspberry Pi 3 is a significant limiting factor for parallel computing on a cluster.
It is not at all suited to computational tasks that require substantial network communications. 

\subsection{Monte Carlo Estimation of Pi}\label{sec:pi}

The second computational task used in examining the performance of our Raspberry Pi compute cluster is
computing an estimation of Pi using Monte Carlo integration~\citep{Neal1993}.  
Specifically, we begin with an implementation of the classic average value method (referred to
as ``method one'' in~\citep{Neal1993}), and implement its obvious parallel extension.

In general, there are much more efficient
ways of estimating Pi.  However, we wanted to use a task relatively simple to explain that is both
easy to parallelize as well as which requires very little communications.  Ultimately, since one desired application
of our cluster is for executing metaheuristics in parallel such as simulated annealing, genetic algorithms, etc,
we also wanted to include a task that involves random number generation since these algorithms rely extensively on
random number generation.  Monte Carlo estimation of Pi fits these requirements well.  The only data transmitted
from the master node to each worker node is the number of iterations to compute, and how many threads to execute
on the worker.  This amounts to two 32 bit integers for each of the seven worker nodes for 448 bits total.
Each worker node responds upon completion with a double value (64 bits for another 448 bits across seven workers), 
which is its estimation of Pi.  This is a total data transmission to and from worker nodes of 896 bits.
The time required to transmit 896 bits is negligible, even for the Raspberry Pi 3's slow networking.

You can compute an estimation of Pi using Monte Carlo simulation as follows:
\begin{equation}
\pi_N = \frac{4}{N} \sum_{i=1}^{N} \sqrt{1-U_i^2} ,
\end{equation}
where the $U_i$ are samples drawn uniformly at random from the interval $[0,1)$, and $N$ is the number of samples.
The larger the value of $N$, the more accurate the estimate.

This is easy to parallelize.  You can distribute the computation over $T$ threads by having each thread compute
$\pi_{N/T}$ and then averaging those results.  It is easy to show this to be equivalent as follows:
\begin{equation}
\frac{1}{T} \sum_1^T \pi_{N/T} = \frac{1}{T} \sum_1^T ( \frac{4}{N/T} \sum_{i=1}^{N/T} \sqrt{1-U_i^2} ) = \frac{1}{T} \frac{4}{N/T} \sum_{i=1}^{N} \sqrt{1-U_i^2} = \pi_N .
\end{equation}

Table~\ref{tab:pi} shows a comparison of single node concurrent execution locally on the master node
versus remote execution on a single worker node, for one to four concurrent threads.  The times are in
seconds and are averages of 10 runs.  The data in the table are for two length runs in total number of 
iterations: $1.2 * 10^8$ and $1.2 * 10^9$.  These run lengths are excessive for the task itself, however,
we wanted to use a run length of sufficient length for later exploring the speedup of distributing the task
across all nodes of the cluster.
You can see that the overhead from the RMI calls for this task,
which requires very little data transmission, is negligible.  The times are nearly identical, with the exception
of the four thread runs for the longer run of $1.2 * 10^9$ iterations, where the remote execution on a worker node
used 10\% more time.  In some cases, we see the peculiar
behavior of the remote case completing in less time than the local case.

\begin{table}[t]
\centering
\caption{Timing results for matrix multiplication on single Raspberry Pi.  Times are in seconds.}\label{tab:pi}
\begin{tabular}{r|rr|rr}\hline
		& \multicolumn{2}{c}{$1.2 * 10^8$ iterations}	 &	\multicolumn{2}{|c}{$1.2 * 10^9$ iterations} \\
Number  &	Local execution &	Remote execution	& Local execution	& Remote execution \\
of threads	& on master node	& on single worker	& on master node	& on single worker \\\hline
1 &	17.98 &	18.04 &	179.72 &	179.54 \\
2 &	9.05 &	9.06 &	89.95 &	89.88 \\
3 &	6.07 &	6.06 &	60.06 &	60.03 \\
4 &	4.58 &	4.58 &	45.10 &	49.29 \\\hline
\end{tabular}
\end{table}

\begin{figure}[t!]
\centering
\includegraphics[scale=0.88]{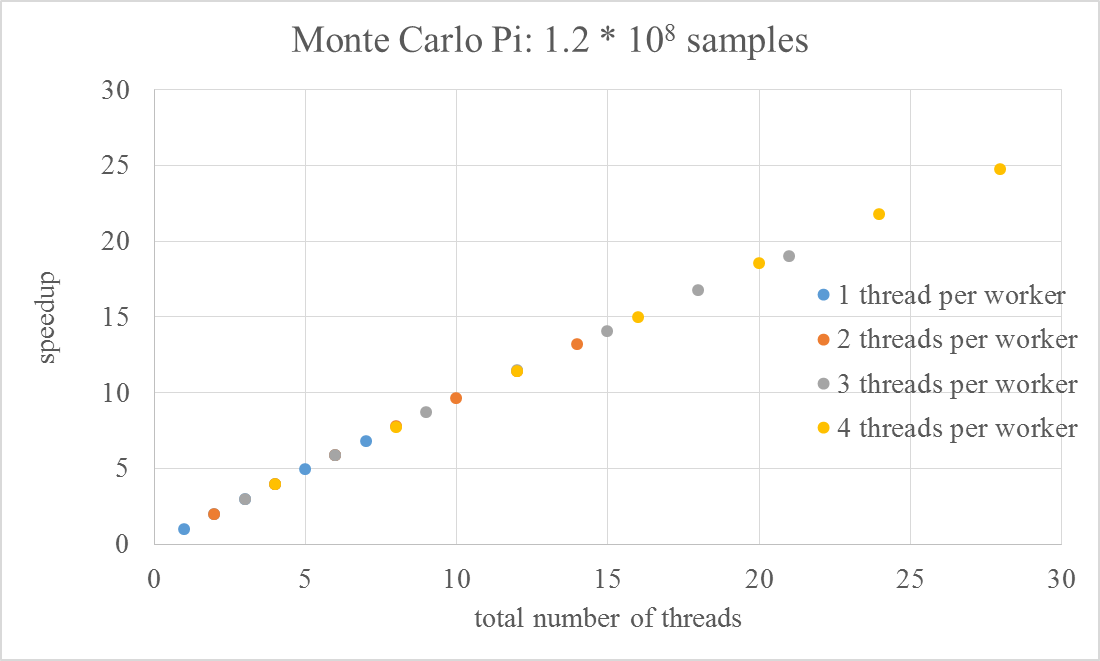}
\caption{Speedup relative to sequential implementation for $1.2 * 10^8$ iterations.}\label{fig:10to8}
\end{figure}

\begin{figure}[t]
\centering
\includegraphics[scale=0.88]{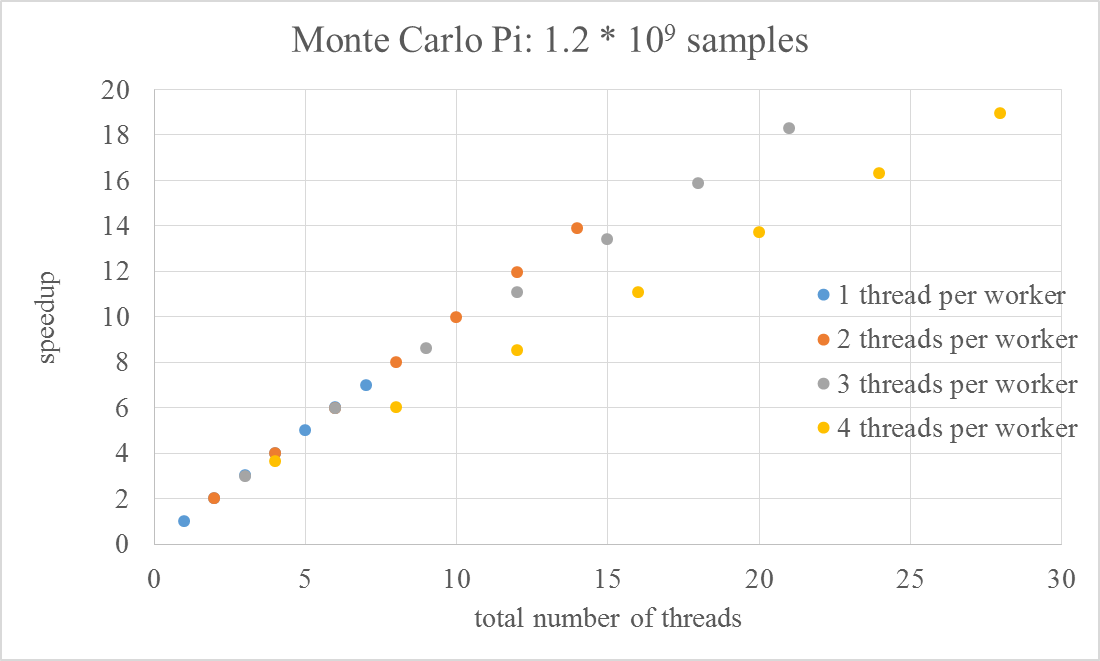}
\caption{Speedup relative to sequential implementation for $1.2 * 10^9$ iterations.}\label{fig:10to9}
\end{figure}

We next consider distributing the computation of the Pi estimation over the entire cluster.
Specifically, we consider from 1 to 7 worker nodes, and from 1 to 4 threads per worker node.
For each of the 28 combinations of number of workers and number of threads per worker, we execute 10 
runs, and average the time to estimate Pi over those 10 runs.  We repeat for two length runs in number
of Monte Carlo samples: $1.2 * 10^8$ and $1.2 * 10^9$.
The results are shown in Figure~\ref{fig:10to8} for the shorter runs, 
and Figure~\ref{fig:10to9} for the longer runs.  These figures show speedup relative to the sequential
implementation.  The x axis in both graphs is total number of threads.

First consider the shorter run length (results of Figure~\ref{fig:10to8}).  Speedup from parallelizing
the computation is approximately linear, regardless of number of threads per worker node, and regardless
of number of worker nodes.

Next, consider the longer run length (results of Figure~\ref{fig:10to9}).  For one or two threads per worker node
(blue and orange in the graph, respectively), speedup is linear.  For three threads per worker (gray in the graph), 
speedup is very nearly linear, with slightly sublinear speedup at five or more worker nodes (15, 18, or 21 total threads).
However, at four threads per worker, speedup is sublinear, with only marginal time improvement from the addition of the
fourth thread per worker.  Given the Raspberry Pi 3 is quad-core, four threads is hitting the concurrency limit of
the processor.  The longer run of the experiment shown in this graph is thus more likely to result in contention for the
processor from operating system level processes, etc.  It is also possible that the longer run with all four processor cores
running continuously may be pushing the Raspberry Pi to the temperature (80 degrees Celsius) 
at which the Pi throttles the CPU clock speed down.  I did not 
record system temperature during the experiment, so I do not know
if this is the case.  However, I don't believe this is likely as I 
would expect the timing results to be further off than they are
if the clock speed was cut in half for part of the run.

Unlike the case of matrix multiplication, we are able to achieve linear speedup from parallelizing Monte Carlo estimation.
The inherent difference between these two tasks that leads to this is that in the case of matrix multiplication, the time
required to distribute the matrix strongly dominates the time required for the actual computation, while in the case of
Monte Carlo simulation, very little data is transmitted.

\section{Conclusion}\label{sec:conclude}

In this report, we describe the implementation of an inexpensive, simple 32 core Raspberry Pi cluster.  
We explore its performance on a couple
easily parallelized computational tasks, one that requires substantial data distribution relative to computational workload, and
one that does not.  Our results show that a cluster of Raspberry Pi single board computers is significantly limited
by its 100 Mbps networking, and is thus not suited to tasks that require distributing a significant amount of data relative
to computation time.  However, we can achieve linear speedup for 
tasks that either require very little data transfer among cluster nodes, or that have relatively long
run lengths relative to data transfer times.  We also observed that speedup from parallelization drops off significantly
at four threads per worker node, and recommend limiting concurrency to three threads per worker to minimize resource
contention with OS processes.

\bibliographystyle{plainnat}
\bibliography{pi}

\begin{thebibliography}{14}
\providecommand{\natexlab}[1]{#1}
\providecommand{\url}[1]{\texttt{#1}}
\expandafter\ifx\csname urlstyle\endcsname\relax
  \providecommand{\doi}[1]{doi: #1}\else
  \providecommand{\doi}{doi: \begingroup \urlstyle{rm}\Url}\fi

\bibitem[Adams et~al.(2015)Adams, Caswell, Matthews, Peck, Shoop, and
  Toth]{Adams:2015}
Joel~C. Adams, Jacob Caswell, Suzanne~J. Matthews, Charles Peck, Elizabeth
  Shoop, and David Toth.
\newblock Budget beowulfs: A showcase of inexpensive clusters for teaching pdc.
\newblock In \emph{Proceedings of the 46th ACM Technical Symposium on Computer
  Science Education}, pages 344--345. ACM, 2015.

\bibitem[Adams et~al.(2016)Adams, Caswell, Matthews, Peck, Shoop, Toth, and
  Wolfer]{Adams:2016}
Joel~C. Adams, Jacob Caswell, Suzanne~J. Matthews, Charles Peck, Elizabeth
  Shoop, David Toth, and James Wolfer.
\newblock The micro-cluster showcase: 7 inexpensive beowulf clusters for
  teaching pdc.
\newblock In \emph{Proceedings of the 47th ACM Technical Symposium on Computing
  Science Education}, pages 82--83. ACM, 2016.

\bibitem[Ballard et~al.(2012)Ballard, Demmel, Holtz, Lipshitz, and
  Schwartz]{Ballard:2012}
Grey Ballard, James Demmel, Olga Holtz, Benjamin Lipshitz, and Oded Schwartz.
\newblock Communication-optimal parallel algorithm for strassen's matrix
  multiplication.
\newblock In \emph{Proceedings of the Twenty-fourth Annual ACM Symposium on
  Parallelism in Algorithms and Architectures}, pages 193--204. ACM, 2012.

\bibitem[Chatterjee et~al.(1999)Chatterjee, Lebeck, Patnala, and
  Thottethodi]{Chatterjee:1999}
Siddhartha Chatterjee, Alvin~R. Lebeck, Praveen~K. Patnala, and Mithuna
  Thottethodi.
\newblock Recursive array layouts and fast parallel matrix multiplication.
\newblock In \emph{Proceedings of the Eleventh Annual ACM Symposium on Parallel
  Algorithms and Architectures}, pages 222--231. ACM, 1999.

\bibitem[Cicirello(2017{\natexlab{a}})]{Cicirello:GitHub:2017}
Vincent~A. Cicirello.
\newblock Performance tests for small clusters.
\newblock GitHub, August 2017{\natexlab{a}}.
\newblock Source code repository:
  \url{https://github.com/cicirello/ClusterPerformanceTests}.

\bibitem[Cicirello(2017{\natexlab{b}})]{cicirello2017SoCS2}
Vincent~A. Cicirello.
\newblock Variable annealing length and parallelism in simulated annealing.
\newblock In \emph{Proceedings of the Tenth International Symposium on
  Combinatorial Search (SoCS 2017)}, pages 2--10. AAAI Press, June
  2017{\natexlab{b}}.
\newblock \doi{10.1609/socs.v8i1.18424}.

\bibitem[Cormen et~al.(2009)Cormen, Leiserson, Rivest, and
  Stein]{cormen2009introduction}
T.H. Cormen, C.E. Leiserson, R.L. Rivest, and C.~Stein.
\newblock \emph{Introduction to Algorithms}.
\newblock MIT Press, 2009.

\bibitem[Hall(1892)]{BeowulfEpic}
Lesslie Hall, editor.
\newblock \emph{Beowulf}.
\newblock D.C. Heath and Co., 1892.
\newblock English translation.

\bibitem[Karstadt and Schwartz(2017)]{Karstadt:2017}
Elaye Karstadt and Oded Schwartz.
\newblock Matrix multiplication, a little faster.
\newblock In \emph{Proceedings of the 29th ACM Symposium on Parallelism in
  Algorithms and Architectures}, pages 101--110. ACM, 2017.

\bibitem[Matthews(2016)]{Matthews:2016}
Suzanne~J. Matthews.
\newblock Teaching with parallella: A first look in an undergraduate parallel
  computing course.
\newblock \emph{Journal of Computing Sciences in Colleges}, 31\penalty0
  (3):\penalty0 18--27, January 2016.

\bibitem[Neal(1993)]{Neal1993}
David Neal.
\newblock Determining sample sizes for monte carlo integration.
\newblock \emph{The College Mathematics Journal}, 24\penalty0 (3):\penalty0
  254--259, 1993.

\bibitem[{Raspberry Pi Foundation}(2017)]{RaspPi}
{Raspberry Pi Foundation}.
\newblock Raspberry pi: Teach, learn, and make with raspberry pi.
\newblock Website, 2017.
\newblock \url{https://www.raspberrypi.org/}.

\bibitem[Sterling et~al.(1995)Sterling, Savarese, Becker, Dorband, Ranawake,
  and Packer]{BeowulfNasa1995}
Thomas~L. Sterling, Daniel Savarese, Donald~J. Becker, John~E. Dorband,
  Udaya~A. Ranawake, and Charles~V. Packer.
\newblock {BEOWULF:} {A} parallel workstation for scientific computation.
\newblock In \emph{Proceedings of the 1995 International Conference on Parallel
  Processing}, pages 11--14, 1995.

\bibitem[Strassen(1969)]{Strassen:1969}
Volker Strassen.
\newblock Gaussian elimination is not optimal.
\newblock \emph{Numer. Math.}, 13\penalty0 (4):\penalty0 354--356, 1969.

\end{thebibliography}

\end{document}